\documentclass[aps,showpacs,twocolumn,twoside,prc]{revtex4-1}

\usepackage{graphicx}  
\usepackage{epstopdf}
\usepackage{amsmath}
\usepackage{amssymb}
\usepackage{bm}
\usepackage{xcolor,xspace}
\usepackage{hyperref}
\hypersetup{colorlinks=True,urlcolor=blue,linkcolor=blue,
citecolor=blue,filecolor=black}
\usepackage{braket}
\usepackage[percent]{overpic}  

\newcommand{\ba}{\begin{eqnarray}}
\newcommand{\ea}{\end{eqnarray}}
\newcommand{\bsub}{\begin{subequations}}
\newcommand{\esub}{\end{subequations}}

\def\ket#1{|#1\rangle}

\begin{document}
\title{Intertwined Quantum Phase Transitions in the Zr Isotopes}
\author{N.~Gavrielov}\email{noam.gavrielov@mail.huji.ac.il}
\author{A.~Leviatan}\email{ami@phys.huji.ac.il}
\affiliation{Racah Institute of Physics, The Hebrew University, 
Jerusalem 91904, Israel}
\author{F.~Iachello}\email{francesco.iachello@yale.edu}
\affiliation{Center for Theoretical Physics, Sloane Physics Laboratory, 
Yale University, New Haven, Connecticut 06520-8120, USA}

\date{\today}  

\begin{abstract}
We explore the situation of intertwined quantum phase transitions
(IQPTs), for which a QPT involving a crossing of two configurations is
accompanied by a shape evolution of each configuration with its own
separate QPT. We demonstrate the relevance of IQPTs to the Zr isotopes,
with such coexisting Type~I and Type~II QPTs,
and ground state shapes changing from spherical to prolate
axially deformed and finally to gamma-unstable.
Evidence for this scenario is provided by a
detailed comparison with experimental data, using a definite symmetry-based
conceptual framework.
\end{abstract}

\maketitle

\section{Introduction}
Quantum Phase transitions (QPT) have in recent years become of great
interest in a variety of fields~\cite{carr}. In particular, they have been
the subject of many investigations in nuclear 
physics~\cite{jolie,cejnar,caprio,iachello}, where they were originally
introduced~\cite{feng,gilmore}. In this field, most of the attention
has been devoted to shape phase transitions in a single configuration,
described by a single Hamiltonian,
\ba
\label{eq:type-I}
\hat{H} = \left( 1-\xi \right) \hat{H}_{1}+\xi \hat{H}_{2} ~,
\ea
where $\xi $ is the control parameter. As $\xi $ changes from $0$ to $1$, 
the symmetry and equilibrium shape of the system change from those of 
$\hat{H}_{1}$ to those of $\hat{H}_{2} $. For sake of clarity, 
we denote these phase transitions Type~I.

A different type of phase transitions occurs when two (or more)
configurations coexist~\cite{Heyde11}.
In this case, the quantum Hamiltonian has a matrix form~\cite{frank}
\ba
\hat{H} &=&
\left [
\begin{array}{cc}
\hat{H}_{A}(\xi^A) & \hat{W}(\omega) \\ 
\hat{W}(\omega) & \hat{H}_{B}(\xi^B)
\end{array}
\right ] ~,\quad
\label{Hmat}
\ea
where the index $A$, $B$ denotes the two configurations 
and $\hat{W}$ denotes their coupling. We call for sake of
clarity these phase transitions Type~II~\cite{frank},
to distinguish them from those
of a single configuration~\cite{Heyde04}.
The two types of QPTs are usually discussed separately
and both have been established in nuclei,
{\it e.g.}, Type~I QPT in the neutron number 90 region
for Nd-Sm-Gd isotopes, and Type~II QPT in nuclei near shell closure,
{\it e.g.}, in the light Pb-Hg isotopes, with strong mixing between the
two configurations. In the present work, we explore a situation where
the two crossing configurations, although coupled, still maintain
individually a pronounced shape evolution with nucleon number. We refer to
such a scenario as intertwined quantum phase transitions (IQPTs) in the
sense that Type~I and Type~II coexist, and show empirical evidence for it
in transitional nuclei, analyzed in a physically transparent symmetry-based
framework.

\section{Algebraic approach to QPTs}
A variety of methods have been used to study shape phase transitions in
nuclei. We prefer here to use algebraic models, in which both Hamiltonians, 
$\hat{H}_{A}$ and $\hat{H}_{B}$, and their coupling, $\hat{W}$, are 
written in terms of the Interacting Boson Model (IBM)~\cite{iac},
with bosons representing valence nucleon pairs counted from the
nearest closed shells.
This provides a simple tractable shell-model-inspired framework,
where global trends 
of structure and symmetries can be clearly identified and 
diversity of observables calculated. Other 
microscopic but computationally demanding approaches
include mean-field methods, both non-relativistic~\cite{robledo}
and relativistic~\cite{vretenar}, and very recently the Monte-Carlo 
shell-model (MCSM) \cite{taka}. In this paper,
we focus on the $_{40}$Zr isotopes and find a complex variety of 
phase transitions both of Type~I and Type~II coexist, thus exemplifying IQPTs.
These isotopes have been very recently the subject of several experimental 
investigations~\cite{chakraborty,Browne15,pietralla,Ansari17,Paul17,
Witt18,Singh18}.
\begin{table*}[t]
\begin{center}
\caption{\label{tab:parameters}
\small
Parameters of the IBM-CM Hamiltonian, Eq.~(\ref{Hmat-int}),
are in MeV and $\chi$ is dimensionless.
The first row of the Table lists the number of neutrons, and 
particle-bosons $(N,N+2)$ or hole-bosons $(\bar N,\bar N+2)$ in the
$(A,B)$ configurations.}
\begin{tabular}{lcccccccccc}
\hline
          & $52(1,3)$ & $54(2,4)$  & $56(3,5)$ &  $58(4,6)$ & $60(5,7)$ & 
$62(6,8)$ & $64(7,9)$ & $66(8,10)$ & $68(\bar7,\bar9)$	& $70(\bar6,\bar8)$ 
\\[2pt]
\hline
$ \epsilon^{(A)}_d $   & 0.7 & 0.8  & 	1.82   & 1.75   & 1.2 &    
1.2 & 1.2 & 1.2 & 1.2 & 1.2 \\
$ \kappa^{(A)}$ & $-0.005$ & $-0.005$ & $-0.005$ & $-0.007$ & 
$-0.006$ & $-0.006$ & $-0.006$ & $-0.006$ & $-0.006$ & $-0.006$ \\
$ \epsilon^{(B)}_d $  & 0.35 & 0.37 & 0.6 & 0.45    & 0.3 &    
0.15 & 0 &  0 & 0 & 0.15\\
$\kappa^{(B)}$ & $ -0.02 $ & $ -0.02 $ & $ -0.015 $ & $ -0.02 $ & $ -0.02 $ & 
$ -0.025 $ & $ -0.0275 $ & $ -0.03 $ & $ -0.0275 $ & $ -0.025 $ \\
$ \kappa^{\prime(B)}$ & 0.01 & 0.01 & 0.01 & 0.01 & 0.0075 & 
0.01 & 0.0125 & 0.0125 & 0.0125 & 0.01 \\
$ \chi $ & $ -0.6 $ & $ -0.6 $ & $ -0.6 $ & $ -0.6 $ & $ -1.0 $ & $ -1.0 $ & 
$ -0.75 $ & $ -0.25 $ & $ -0.25 $ & 0 \\
$ \Delta^{(B)}_p $ & 1.6 & 1.6 & 1.84 & 1.43 & 0.8 & 
0.8 & 0.8 & 0.8 & 0.8 & 0.8\\
$ \omega $ & 0.1 & 0.1 & 0.02 & 0.02 & 0.02 & 0.02 & 0.02 & 
0.02 & 0.02 & 0.02 \\
\hline 
\end{tabular}
\end{center}
\end{table*}

To be specific, we use the configuration mixing model
(IBM-CM) of~\cite{duval}, and
write the Hamiltonian not in matrix form, but rather in the equivalent form
\ba
\hat{H}=\hat{H}_{A}^{(N)}+\hat{H}_{B}^{(N+2)}+\hat{W}^{(N,N+2)} ~,
\ea
where $\hat{\cal O}^{(N)}=\hat{P}_{N}^{\dag }\hat{\cal O}\hat{P}_{N}$ 
and $\hat{\cal O}^{(N,N^{\prime })}=
\hat{P}_{N}^{\dag }\hat{\cal O}\hat{P}_{N^{\prime }}$, 
for an operator $\hat{\cal O}$, with $\hat{P}_{N}$, a projection operator 
onto the $[N] $ boson space. Here 
$\hat{H}_{A}^{(N)}$ represents the so-called normal 
($N$ boson space) configuration and $\hat{H}_{B}^{(N+2)}$
represents the so-called intruder ($N\!+\!2$ boson space)
configuration, which we have assumed, 
as in~\cite{sambataro} where a similar calculation was done for the 
$_{42}$Mo isotopes, to be a proton excitation across the subshell closure 
at proton number 40 (see Fig.~1 of~\cite{sambataro}). 
The explicit form of these Hamiltonians is 
\begin{subequations}
\label{Hmat-int}
\begin{align}
\hat{H}_{A} =&\,\epsilon _{d}^{(A)}\hat{n}_{d}
+\kappa ^{(A)}\hat{Q}_{\chi}\cdot \hat{Q}_{\chi } ~,
\label{HA}
\\
\hat{H}_{B} =&\,\epsilon _{d}^{(B)}\hat{n}_{d} 
+\kappa ^{(B)}\hat{Q}_{\chi}\cdot \hat{Q}_{\chi} 
+\kappa ^{\prime (B)}\hat{L}\cdot \hat{L} 
+ \Delta_p^{(B)} ~, 
\label{HB}
\end{align}
\end{subequations}
where the quadrupole operator is defined as 
$\hat{Q}_{\chi} = d^{\dag}s+s^{\dag }\tilde{d}
+\chi ( d^{\dag}\times \tilde{d}) ^{(2)}$. 
In Eq.~(\ref{HB}), $\Delta_p^{(B)}$ is the off-set 
between the normal and intruder configurations, where the index $p$ denotes
the fact that this is a proton excitation. The mixing term has the 
form~\cite{iac,duval}
\ba
\label{eq:mixing}
\hat{W} = [\,\omega_d\,( d^{\dag }\times d^{\dag })^{(0)}
+ \,\omega_s\,(s^{\dag })^2\,] + {\rm H.c.} ~,\quad
\ea
where, for simplicity, a single parameter $\omega=\omega_{s}=\omega_{d}$ 
is used. Hamiltonians of the above form, have been used extensively
for studying coexistence phenomena in nuclei~\cite{duval, sambataro,
duval83,Bijker06,Foisson03,ramos11,ramos14,nomurajpg,nomura16,lev18}.
The resulting eigenstates $\ket{\Psi;L}$ 
with angular momentum $L$, are linear combinations of the
wave functions, $\Psi_A$ and $\Psi_B$,
in the two spaces $[N]$ and $[N+2]$,
\ba
\label{wf}
\ket{\Psi; L} = a\ket{\Psi_A; [N], L} + b\ket{\Psi_B; [N+2], L} ~,
\ea
with $a^{2}+b^{2}=1$. We note here that one of the advantages of the
algebraic method is that one can also study phase transitions
semi-classically by introducing intrinsic states~\cite{gin80,diep80}
and constructing the
corresponding energy functional (or potential function).
For a single configuration, the latter is
a scalar function of the quadrupole variables,
$\beta$ and $\gamma$~\cite{iac}.
When two configurations coexist,
the energy functional becomes a matrix. Diagonalization of this two-by-two
matrix produces the so-called eigen-potentials,
$E_{\pm}(\beta,\gamma)$~\cite{frank,frank04,hellemans07}.

\section{QPTs in the Zirconium chain}
\begin{figure}[t]
\begin{overpic}[width=1\linewidth]{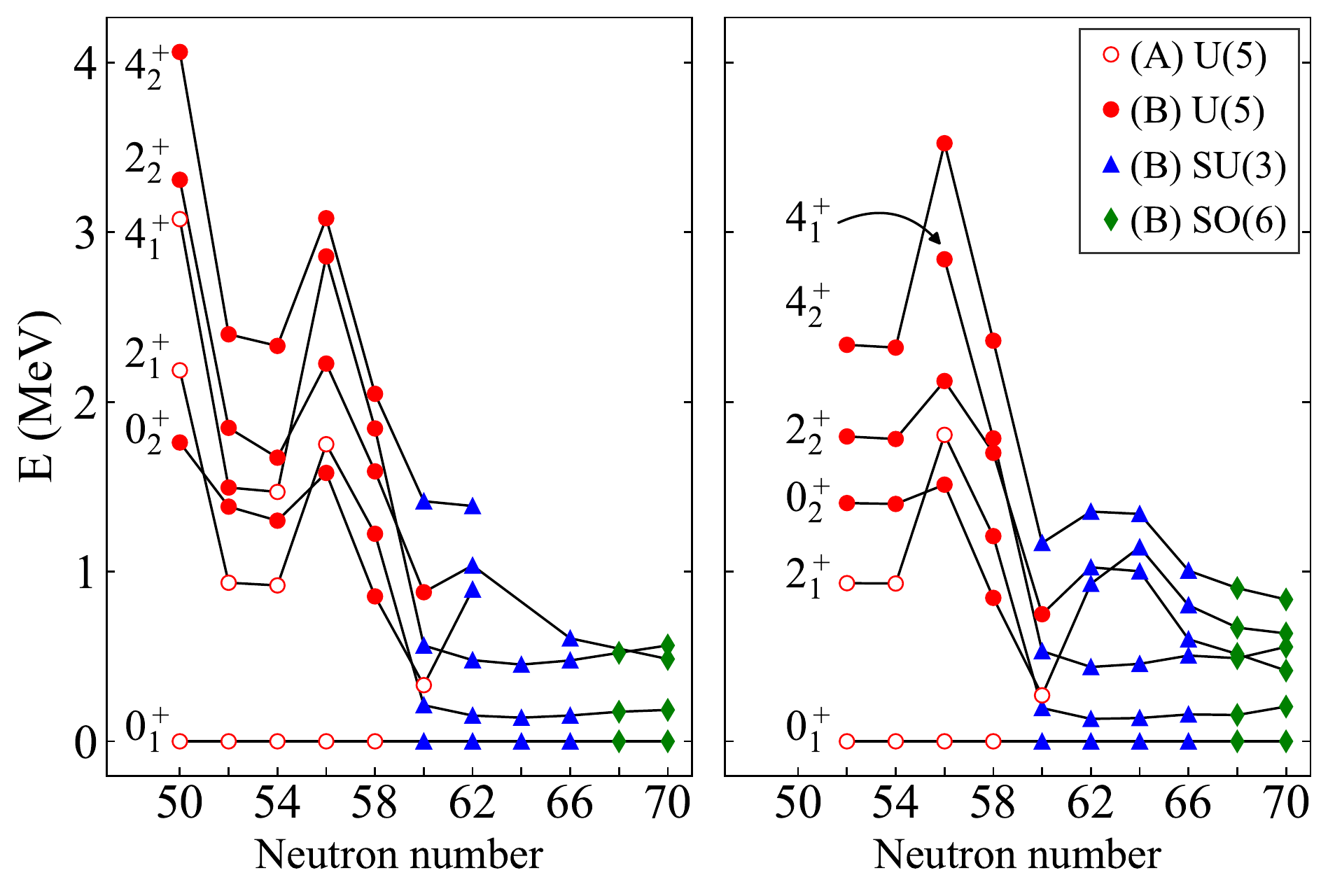}
\put (34,61) {\large (a) {\bf Exp}}
\put (56,61) {\large (b) {\bf Calc}}
\end{overpic}
\caption{Comparison between (a)~experimental~\cite{Paul17,ensdf} and
(b)~calculated energy levels
$0_{1}^{+},2_{1}^{+},4_{1}^{+},0_{2}^{+},2_{2}^{+},4_{2}^{+}$.
Empty (filled) symbols indicate a state dominated by the 
normal $A$-configuration (intruder $B$-configuration),
with assignments based on the decomposition of Eq.~(\ref{wf}).
Note that the calculated values
start at neutron number 52, while the experimental values include the
closed shell at 50.
\label{fig:levels}}
\end{figure}

The IBM-CM framework described above has been previously
employed to the Zr chain in~\cite{nomura16}, where the
Hamiltonian parameters were determined by a mapping between
microscopic-derived and IBM energy surfaces.
Due to the mean-field nature of this procedure, 
the Hamiltonian obtained exhibited noticeable deviations from the data
in the vicinity of the critical point. In the current study,
we adapt a different approach as
in~\cite{sambataro,duval83,Bijker06,Foisson03,ramos11,ramos14}, with
parameters determined from a combined fit to the data on spectra and
$E2$ transitions for the states of $^{92-110}$Zr shown in
Fig.~\ref{fig:levels}, allowing a gradual change between adjacent isotopes,
but taking into account the proposed shell-model interpretation for the
structure evolution in this  region~\cite{Talmi65,FedPit79,HeyCas85}.
The Hamiltonian parameters used are given in 
Table~\ref{tab:parameters} and are consistent with those of previous
calculations in this mass region~\cite{sambataro,duval83,Bijker06},
where a similar fit procedure was employed.
It should be noted that beyond the middle of the 
shell at neutron number 66, bosons are replaced by boson holes~\cite{iac}, 
and denoted by a bar over their number,
and a symmetry about mid-shell was imposed on all parameters
(except~$\chi$), in accord with
 microscopic aspects of the IBM~\cite{IacTal87}.
Apart from some fluctuations due to 
the subshell closure at neutron number 56
(the filling of the $2d_{5/2}$ orbital~\cite{Talmi65}),
the values of the parameters are 
a smooth function of neutron number and, in some cases, a constant.
A notable exception is the sharp decrease by 1~MeV of
the energy off-set parameter $\Delta_p^{(B)}$
beyond neutron number 56. Such a behavior was observed for the
Mo and Ge chains~\cite{sambataro,duval83,Bijker06} and, 
as noted in~\cite{sambataro}, it reflects the effects of the isoscalar
residual interaction, $V_{pn}$, between protons and neutrons occupying the
partner orbitals $1g_{9/2}$ and $1g_{7/2}$,
which is the established mechanism for descending cross shell-gap
excitations and onset of deformation in this region~\cite{FedPit79,HeyCas85}.
This trend in $\Delta_p^{(B)}$ agrees with shell model estimates for the
monopole correction of $V_{pn}$~\cite{Heyde87}.
The parameter $\omega$~(\ref{eq:mixing}) is determined from
$E2$ transitions in-between configurations, and is constant
except for $^{92,94}$Zr where the normal configuration space
is small ($N\!=\!1,2$).
Fine-tuning the parameters for
individual isotopes can improve the fit, but the main conclusions of the
analysis, to be reported below, are not changed.
\begin{figure}[t!]
\centering
\includegraphics[width=1\linewidth]{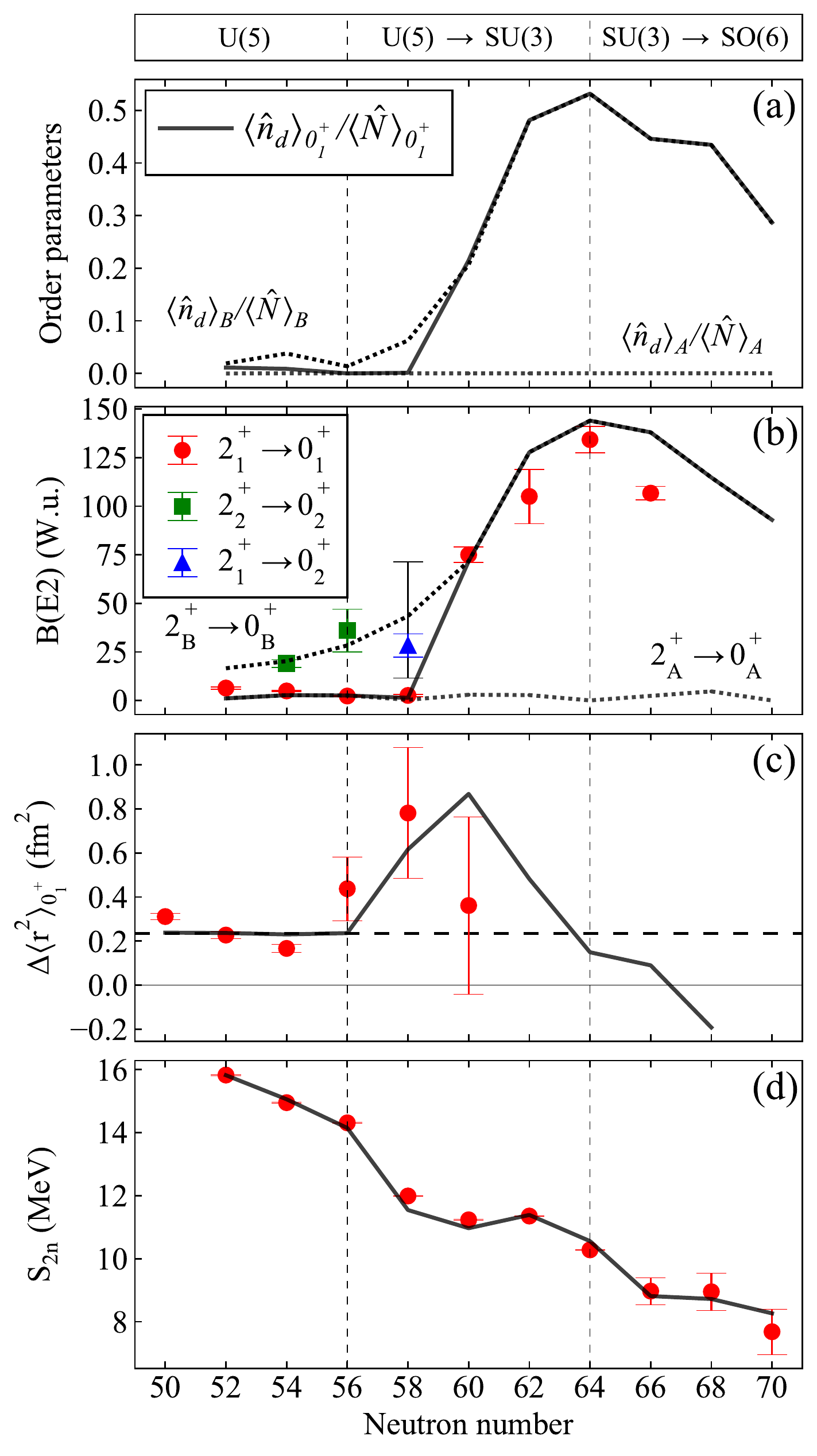}
\caption{Evolution of order parameters and of observables along the Zr chain.
Symbols (solid lines) denote experimental data (calculated results).
Relevant parameters are given in the text.
(a)~Order parameters (see text for details).
(b)~$B(E2)$ values in Weisskopf units (W.u.).
Data taken from~\cite{ensdf,chakraborty,Browne15,pietralla,
Ansari17,Witt18,Singh18}. 
Dotted lines denote calculated $E2$ transitions within a configuration.
(c)~Isotope shift, $\Delta\braket{\hat{r}^{2}}_{0^{+}_1}$ 
in fm$^{2}$. Data taken from~\cite{angeli2}. The horizontal 
dashed line at $0.235$ fm$^{2}$ represents the smooth behavior 
in $\Delta \braket{\hat{r}^{2}}_{0^{+}_1}$ due to the $A^{1/3}$ increase 
of the nuclear radius. 
(d)~Two-neutron separation energies, $S_{2n}$, in MeV. 
Data taken from AME2016~\cite{wang-masses}.
\label{fig:fig-combined}}
\end{figure}
\begin{figure*}[t]
\centering
\begin{overpic}[width=0.28\linewidth]{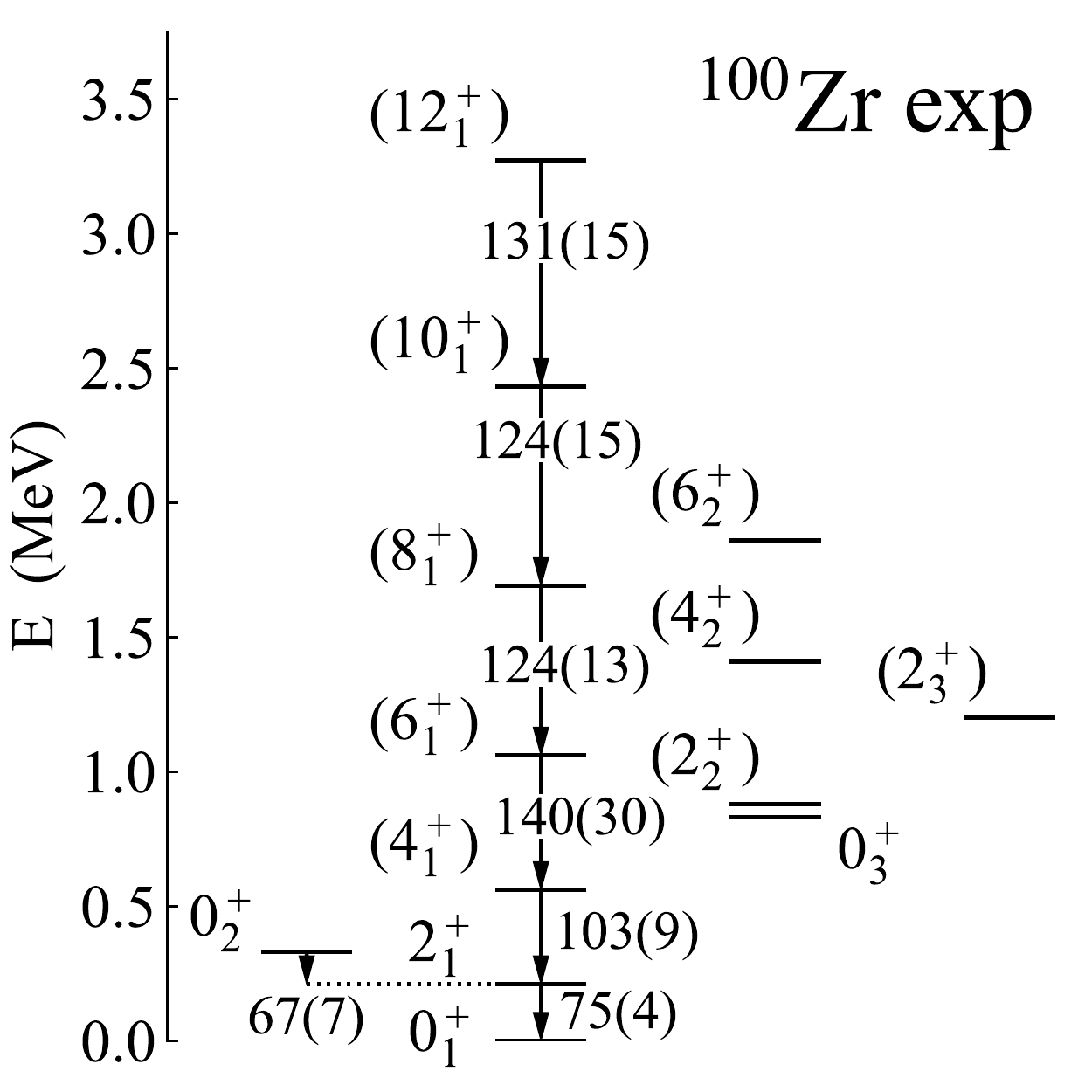}
\put (85,75) {\large(a)}
\end{overpic}
\begin{overpic}[width=0.28\linewidth]{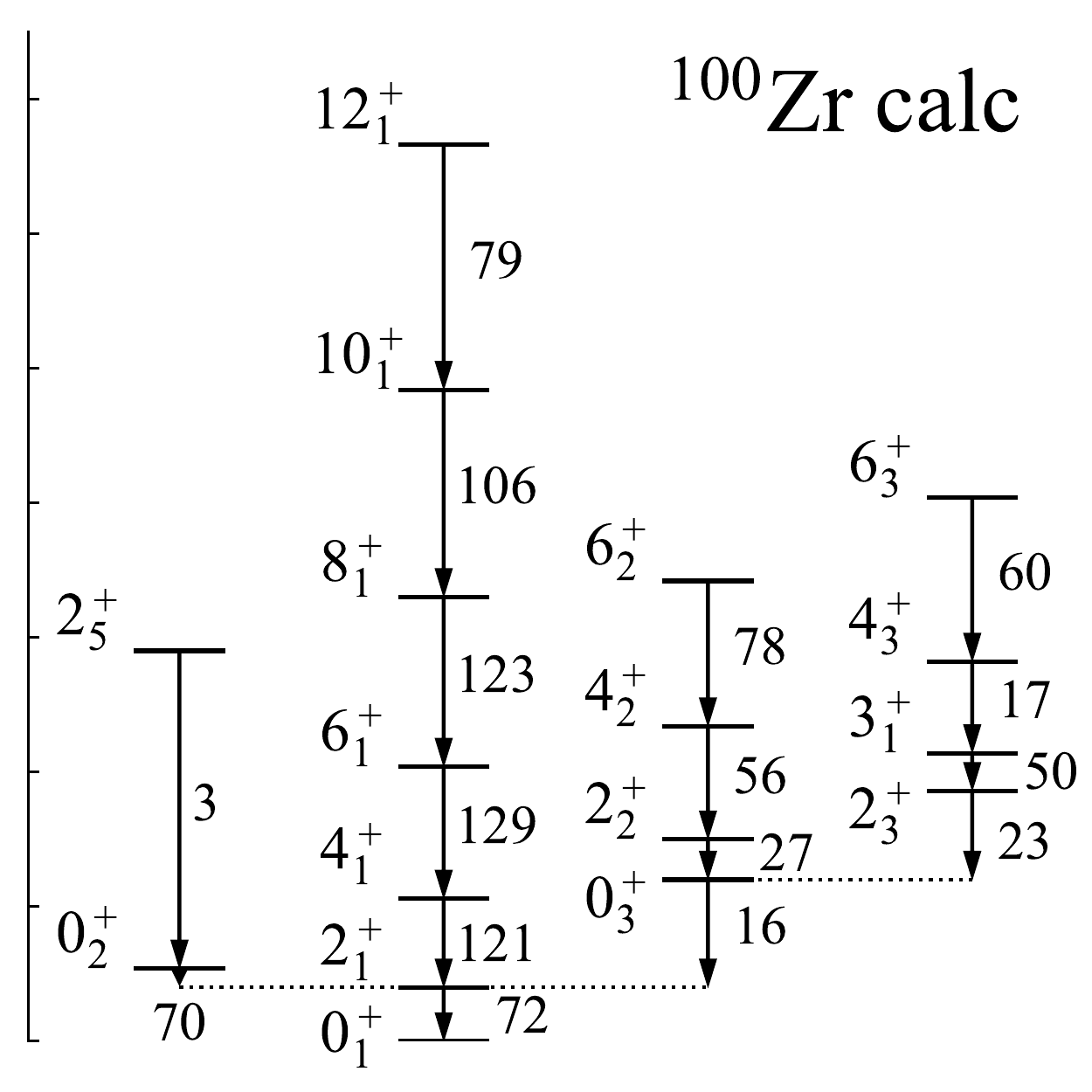}
\put (85,75) {\large(b)}
\end{overpic}
\begin{overpic}[width=0.14\linewidth]{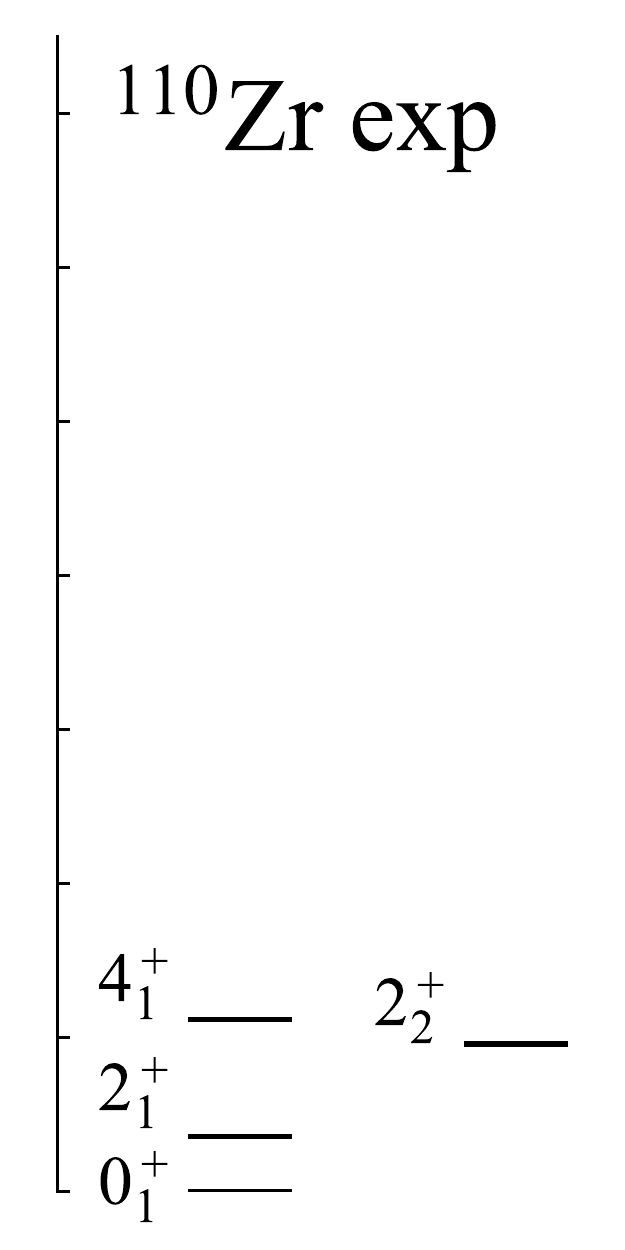}
\put (30,75) {\large(c)}
\end{overpic}
\begin{overpic}[width=0.28\linewidth]{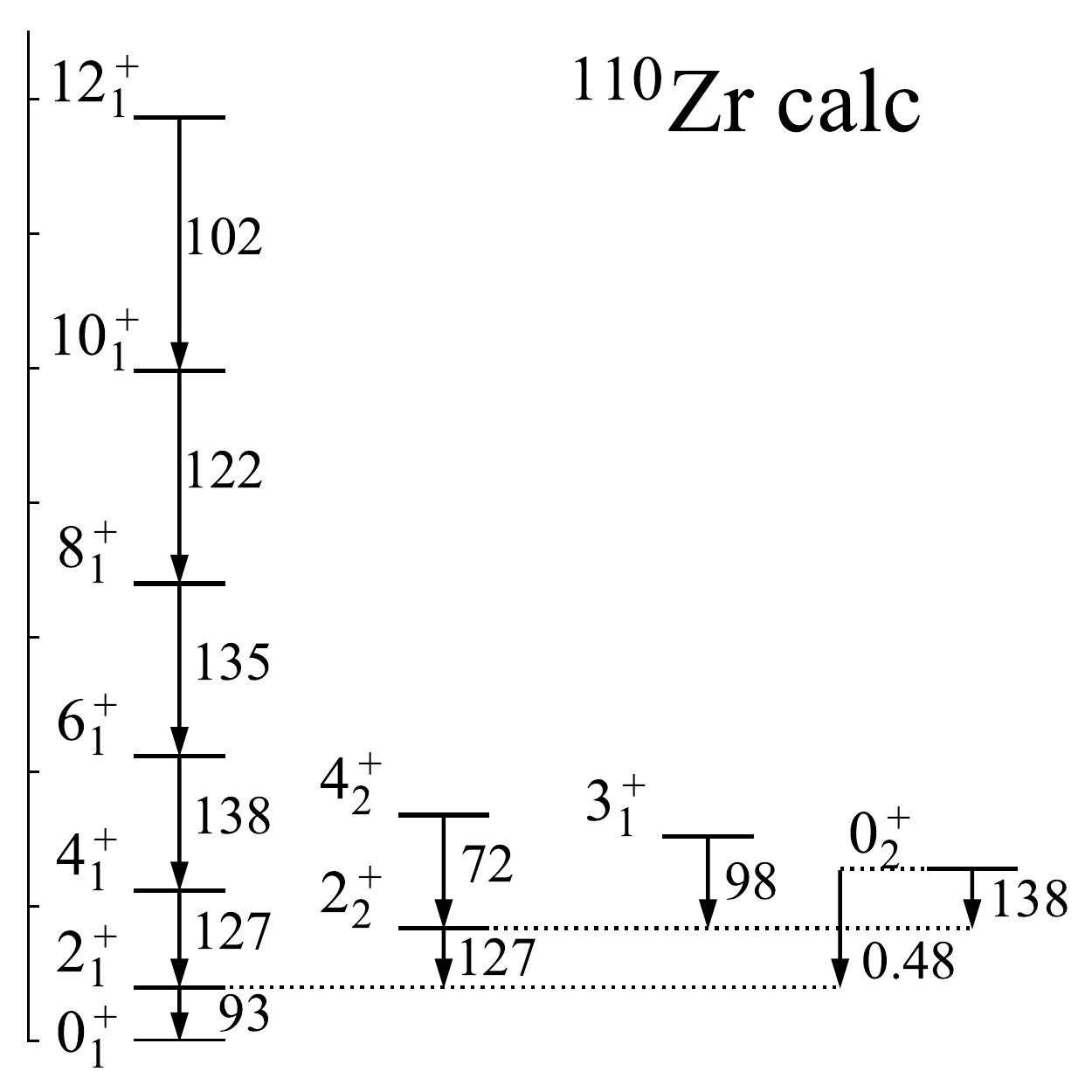}
\put (75,75) {\large(d)}
\end{overpic}
\caption{\label{fig:spectrum}
\small
Experimental and calculated energy levels in MeV 
and $E2$ rates in W.u. for $^{100}$Zr [panels (a)-(b)],
and $^{110}$Zr [panels (c)-(d)].}
\end{figure*}

In Fig.~\ref{fig:levels} we show a comparison between experimental and 
calculated levels. One can see here a rather complex structure.
In the region between neutron number 50 and 56, there appear to be two
configurations, one spherical (seniority-like), ($A$), and one weakly
deformed, ($B$), as evidenced by the ratio $R_{4/2}$, which is at 52-56, 
$R^{(A)}_{4/2}\cong 1.6 $ and  $R^{(B)}_{4/2} \cong 2.3$. 
From neutron number 58, there is a pronounced drop in energy
for the states of configuration~$B$ and at 60, the two configurations
exchange their role indicating a Type~II QPT. 
At this stage,
the intruder configuration~($B$) appears to be at the critical point of
a U(5)-SU(3) Type~I QPT, as evidenced by the low value of the excitation
energy of the first excited $ 0^+ $ state of this configuration
(the $0^{+}_3$ state in $^{100}$Zr shown in Fig.~\ref{fig:spectrum}).
The same situation is seen in the $_{62}$Sm
and $_{64}$Gd isotopes at neutron number 90~\cite{iac,scholten78}.
Beyond neutron number 60, the intruder configuration~($B$) is strongly 
deformed, as evidenced by the small value of the
excitation energy of the state $2_{1}^{+}$, $E_{2_{1}^{+}}\!=\!139.3$ keV and by
the ratio $R^{(B)}_{4/2}\!=\!3.24$ in $^{104}$Zr. At still larger
neutron number 66,
the ground state band becomes $\gamma $-unstable (or triaxial) as evidenced
by the close energy of the states $2_{2}^{+}$ and $4_{1}^{+}$, 
$E_{2_{2}^{+}}\!=\!607.0$~keV, $E_{4_{1}^{+}}\!=\!476.5$ keV, in $^{106}$Zr, 
and especially by the recent results 
$ E_{4^+_1}\!=\!565$~keV and $ E_{2^+_2}\!=\!485$ keV 
in $^{110} $Zr \cite{Paul17}, a signature of the SO(6) symmetry. 
In this region, the ground state 
configuration undergoes a~crossover from SU(3) to SO(6). 

The above spectral analysis suggests a situation of coexisting
Type~I and Type~II QPTs, which is the defining property of IQPTs.
In order to understand the nature of these phase transitions, 
one needs to study the behavior of the order parameters. 
In the present study, the latter involve 
the expectation value of $\hat{n}_d$ in the ground state wave function,
$\ket{\Psi; L\!=\!0^{+}_1}$ and in its
$\Psi_A$ and $\Psi_B$ components~(\ref{wf}), denoted by 
$\braket{\hat{n}_d}_{0^{+}_1}$, $\braket{\hat{n}_d}_A$,
$\braket{\hat{n}_d}_B$, respectively.
$\braket{\hat{n}_d}_A$ and $\braket{\hat{n}_d}_B$ portray 
the shape-evolution in configuration~($A$) and ($B$), respectively, 
and $\braket{\hat{n}_d}_{0^{+}_1}= a^2\braket{\hat{n}_d}_A 
+ b^2\braket{\hat{n}_d}_B$
contains information on the normal-intruder mixing.
Fig.~\ref{fig:fig-combined}(a) shows the evolution along the Zr chain 
of these order parameters 
($\braket{\hat{n}_d}_{A},\,\braket{\hat{n}_d}_{B}$ in dotted 
and $\braket{\hat{n}_d}_{0^{+}_1}$ in solid lines),
normalized by the respective boson numbers,
$\braket{\hat{N}}_A\!=\!N$, 
$\braket{\hat{N}}_B\!=\!N\!+\!2$,
$\braket{\hat{N}}_{0^{+}_1}\!=\!a^2N\!+\!b^2(N\!+\!2)$. 
Configuration~($A$) is seen to be spherical for all
neutron numbers considered. 
In contrast, configuration~($B$) is weakly-deformed
for neutron number 52-58. One can see here clearly a jump 
between neutron number 58 and 60 from
configuration~($A$) to configuration~($B$), indicating a 
1$^{st}$ order Type~II phase transition~\cite{frank}, 
a further increase at neutron numbers 60-64
indicating a U(5)-SU(3) Type~I phase transition, and, finally, 
there is a decrease at neutron number 66, due in part to the 
crossover from SU(3) to SO(6) and in part to the shift from boson 
particles to boson holes after the middle of the major shell 50-82. 
$\braket{\hat{n}_d}_{0^{+}_1}$ is close to $\braket{\hat{n}_d}_A$
for neutron number 52-58 and coincides with $\braket{\hat{n}_d}_B$ 
at 60 and above, consistent with a high degree of purity with respect 
to configuration-mixing.
These conclusions are stressed by an analysis of other observables, 
in particular, the $B(E2)$ values.
Adapted to two configurations, the $E2$ operator reads 
$\hat{T}(E2)\!=\!e^{(A)}\hat Q^{(N)}_{\chi}+e^{(B)}\hat Q^{(N+2)}_{\chi}$, 
with $\hat{Q}_{\chi}^{(N)}\!=\!\hat{P}_{N}^{\dag }\hat{Q}_{\chi}\hat{P}_{N}$
and $\hat{Q}_{\chi}^{(N+2)}\!=\!P_{N+2}^{\dag }\hat{Q}_{\chi}\hat{P}_{N+2}$.
The boson effective charges $e^{(A)}\!=\!0.9$ and
$e^{(B)}\!=\!2.24$ $({\rm W.u.})^{1/2}$ are determined from
the $2^{+}\!\to\! 0^{+}$ transition within each configuration,
and $\chi$ is the same parameter as in the Hamiltonian~(\ref{Hmat-int}).
As shown in Fig.~\ref{fig:fig-combined}(b), the calculated $B(E2)$'s 
agree with the empirical values and follow the same trends as the 
respective order parameters.

Further evidence can be obtained from an analysis of the isotope shift 
$\Delta\braket{\hat r^2}_{0^+_1}
=\braket{\hat{r}^{2}}_{0^{+}_1;A+2}-\braket{\hat{r}^2}_{0^{+}_1;A}$, where 
$\braket{\hat r^2}_{0^+_1} $ is the expectation value of $ \hat r^2 $ in the 
ground state $ 0^+_1 $. In the IBM-CM the latter is given by 
$\braket{\hat r^2} = r^2_c + \alpha N_v + \eta [\braket{\hat n_d^{(N)}} 
+ \braket{\hat n_d^{(N+2)}}]$, 
where $r^2_c$ is the square radius of the closed shell, 
$N_v$ is half the number of valence particles, 
and $\eta$ is a coefficient that takes into account the effect 
of deformation~\cite{iac,vanisacker}.
$\Delta\braket{\hat r^2}_{0^+_1}$ depends on two parameters,
$\alpha\!=\!0.235,\,\eta \!=\! 0.264$ fm$^2$, whose values are fixed
by the procedure of Ref~\cite{vanisacker}.
$\Delta\braket{\hat r^2}_{0^+_1}$ should increase at the transition
point and decrease and, as seen in Fig.~\ref{fig:fig-combined}(c), 
it does so, although the error bars are large and no data are available
beyond neutron number 60. (In the large $N$ limit, this quantity, 
proportional to the derivative of the order parameter 
$\braket{\hat{n}_d}_{0^{+}_1}$,
diverges at the critical point).
Similarly, the two-neutron separation energies $S_{2n} $ 
can be written as~\cite{iac}, 
$S_{2n} = -\tilde{A} -\tilde{B} N_v \pm S^{\text{def}}_{2n} - \Delta_n$, 
where $S^{\text{def}}_{2n}$ is the contribution of the deformation,
obtained by the expectation value of the Hamiltonian in the
ground state~$ 0^+_1$.
The $ + $ sign applies to particles and the $ - $ sign to holes,
and $\Delta_n $ takes into account the neutron subshell closure at 56, 
$\Delta_n = 0 $ for 50-56 and $ \Delta_n = 2 $ MeV for 58-70.
The value of $ \Delta_n $ is taken from Table XII of \cite{barea} 
and $ \tilde{A}\!=\!-16.5,\,\tilde{B}\!=\!0.758$ MeV are determined
by a fit to binding energies of $^{92,94,96}$Zr.
The calculated $ S_{2n}$, shown in Fig. \ref{fig:fig-combined}(d), 
displays a complex behavior. Between neutron number 52 and~56 
it is a straight line, as the ground state is spherical (seniority-like)
configuration~($A$). After 56, it first goes down due to the subshell closure 
at~56, then it flattens as expected from a 1$^{st}$ order Type~I QPT 
(see, for example the same situation in the $_{62}$Sm
isotopes~\cite{scholten78}). After 62, 
it goes down again due to the increasing of deformation and finally it
flattens as expected from a crossover from SU(3) to SO(6).
\begin{figure*}[]
\centering
\begin{overpic}[width=0.19\linewidth]{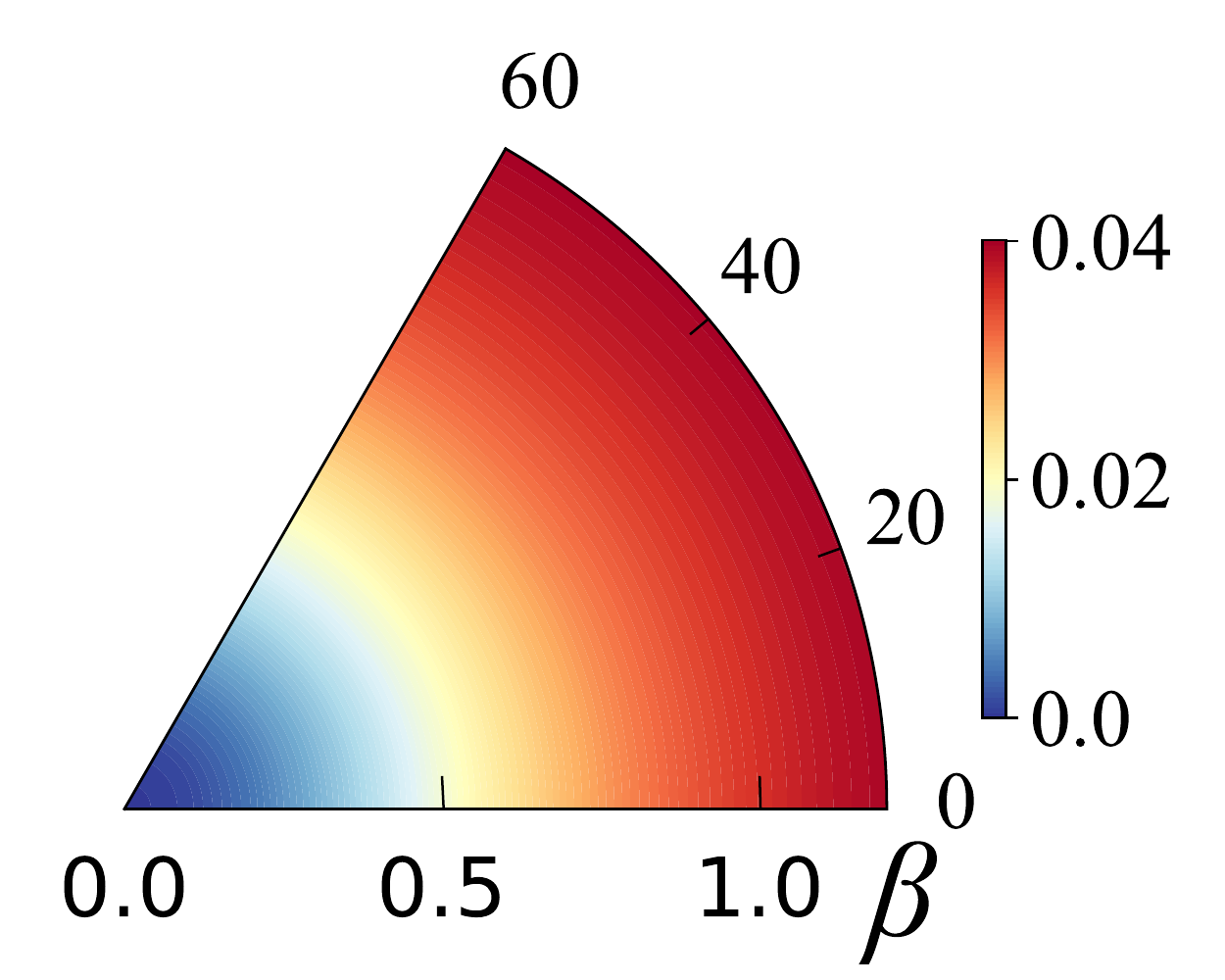}
\put (0,67.5) {\large(a)}
\put (0,50) {\large $^{92}$Zr}
\put (60,70) {$\gamma$(deg)}
\end{overpic}
\begin{overpic}[width=0.19\linewidth]{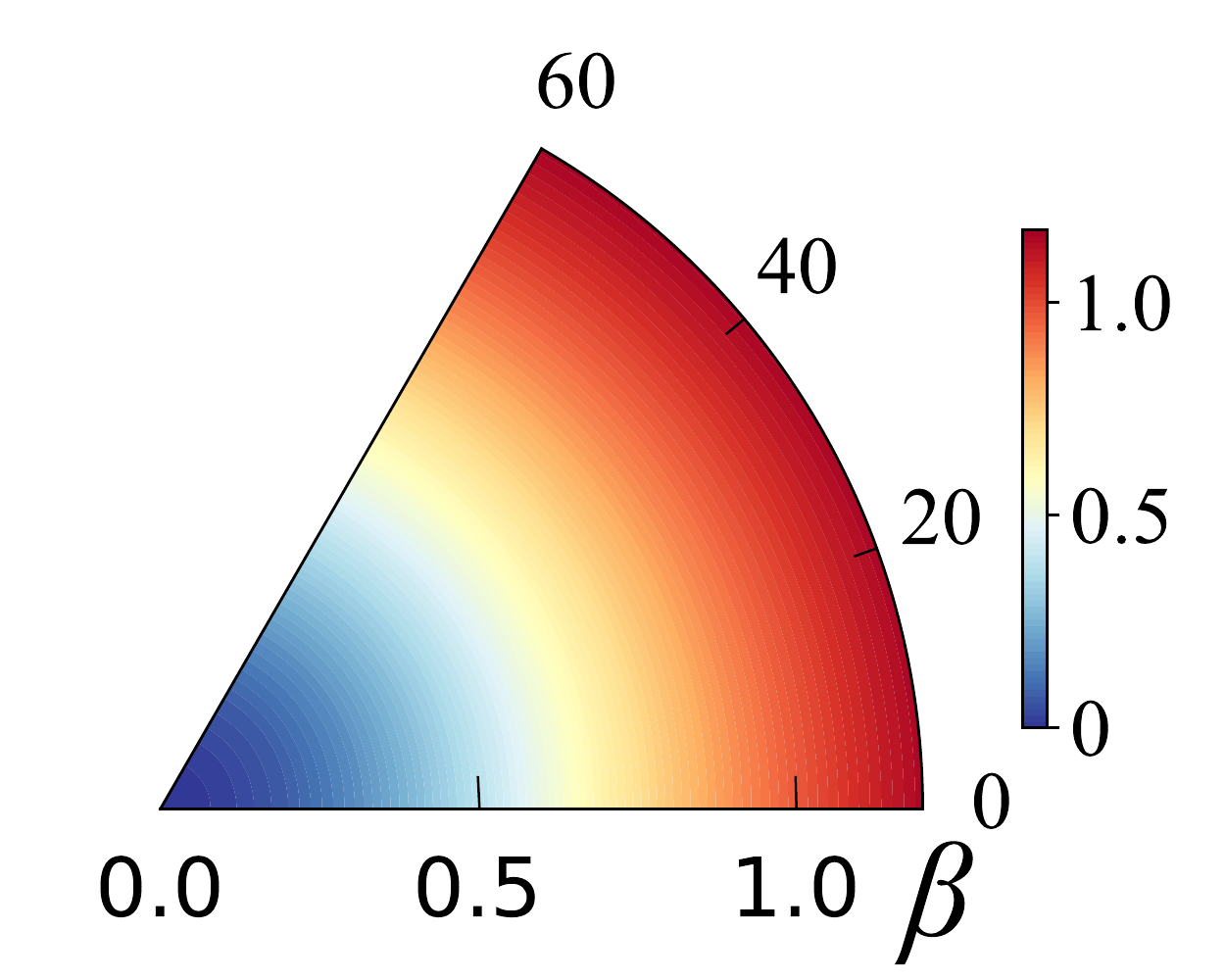}
\put (0,67.5) {\large(b)}
\put (0,50) {\large $^{94}$Zr}
\put (60,70) {$\gamma$(deg)}
\end{overpic}
\begin{overpic}[width=0.19\linewidth]{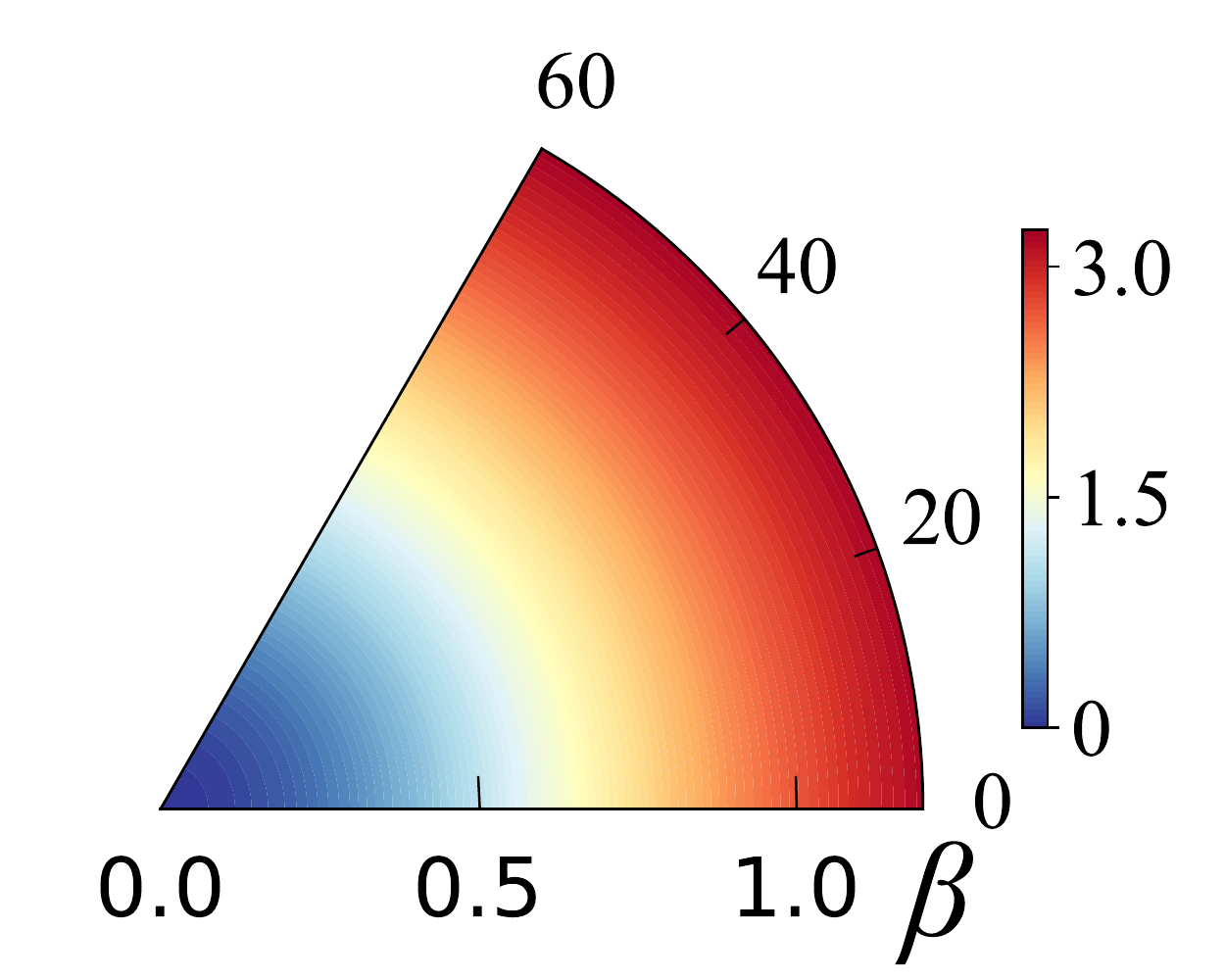}
\put (0,67.5) {\large(c)}
\put (0,50) {\large $^{96}$Zr}
\put (60,70) {$\gamma$(deg)}
\end{overpic}
\begin{overpic}[width=0.19\linewidth]{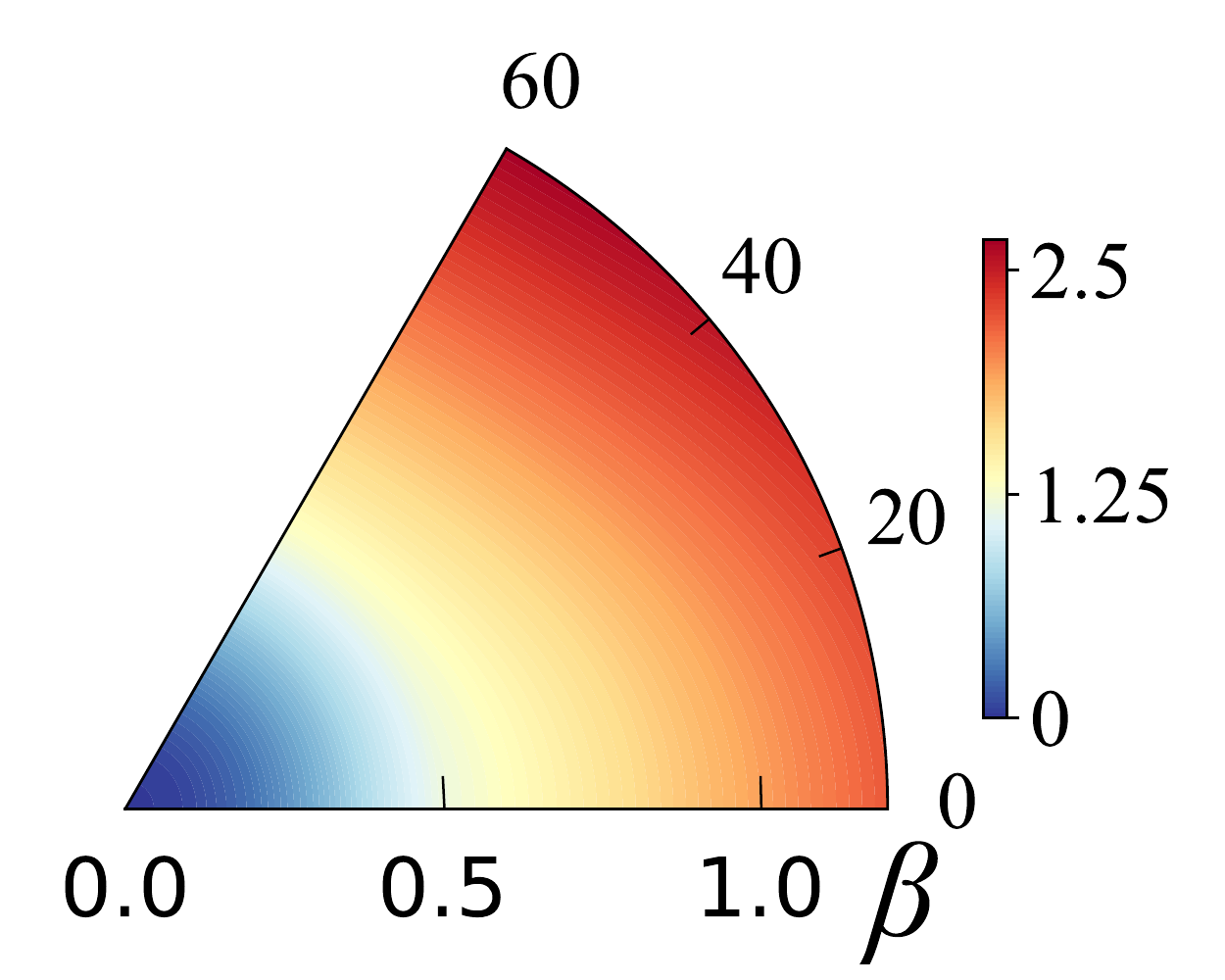}
\put (0,67.5) {\large(d)}
\put (0,50) {\large $^{98}$Zr}
\put (60,70) {$\gamma$(deg)}
\end{overpic}
\begin{overpic}[width=0.19\linewidth]{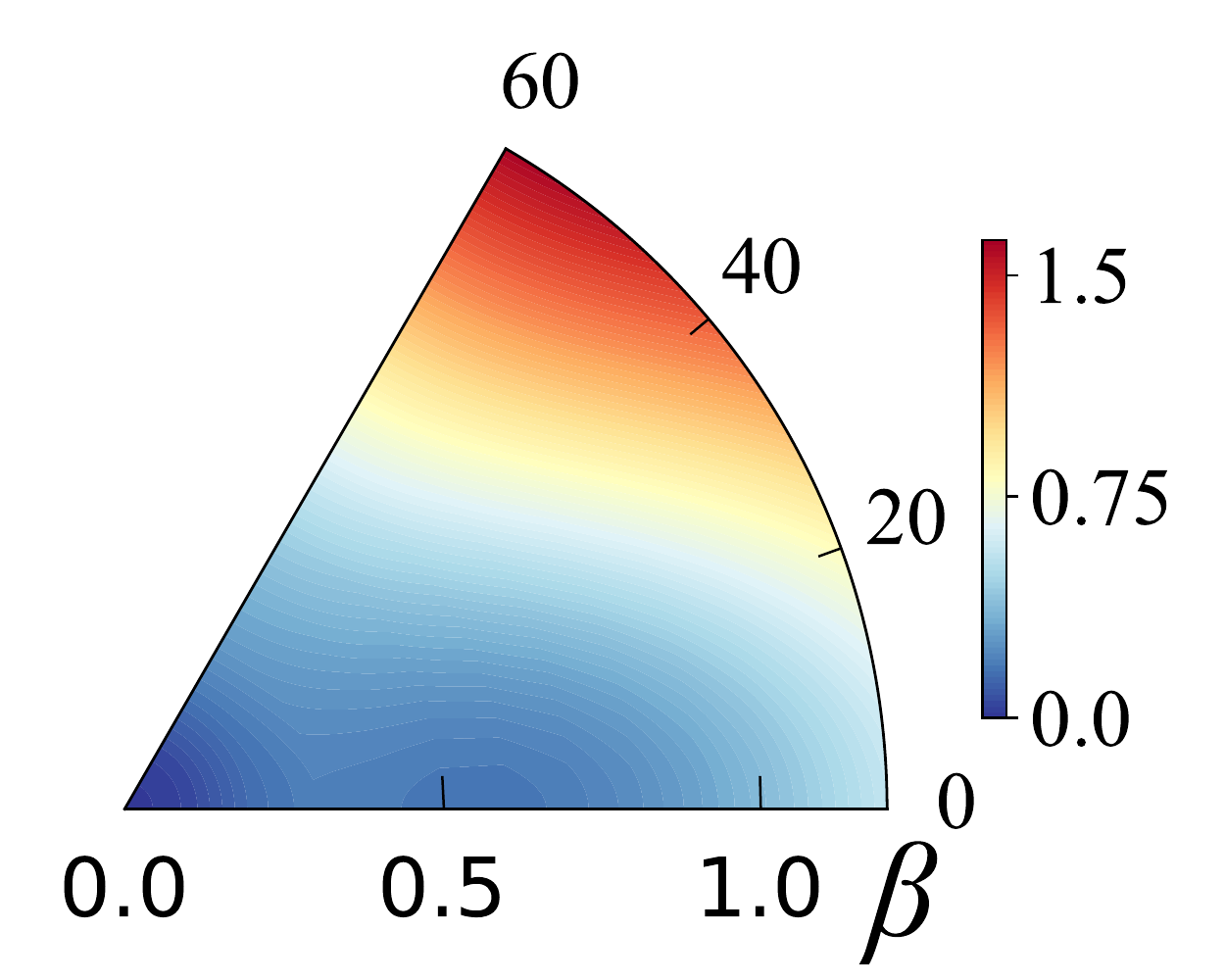}
\put (0,67.5) {\large(e)}
\put (0,50) {\large $^{100}$Zr}
\put (60,70) {$\gamma$(deg)}
\end{overpic} \\ 
\begin{overpic}[width=0.19\linewidth]{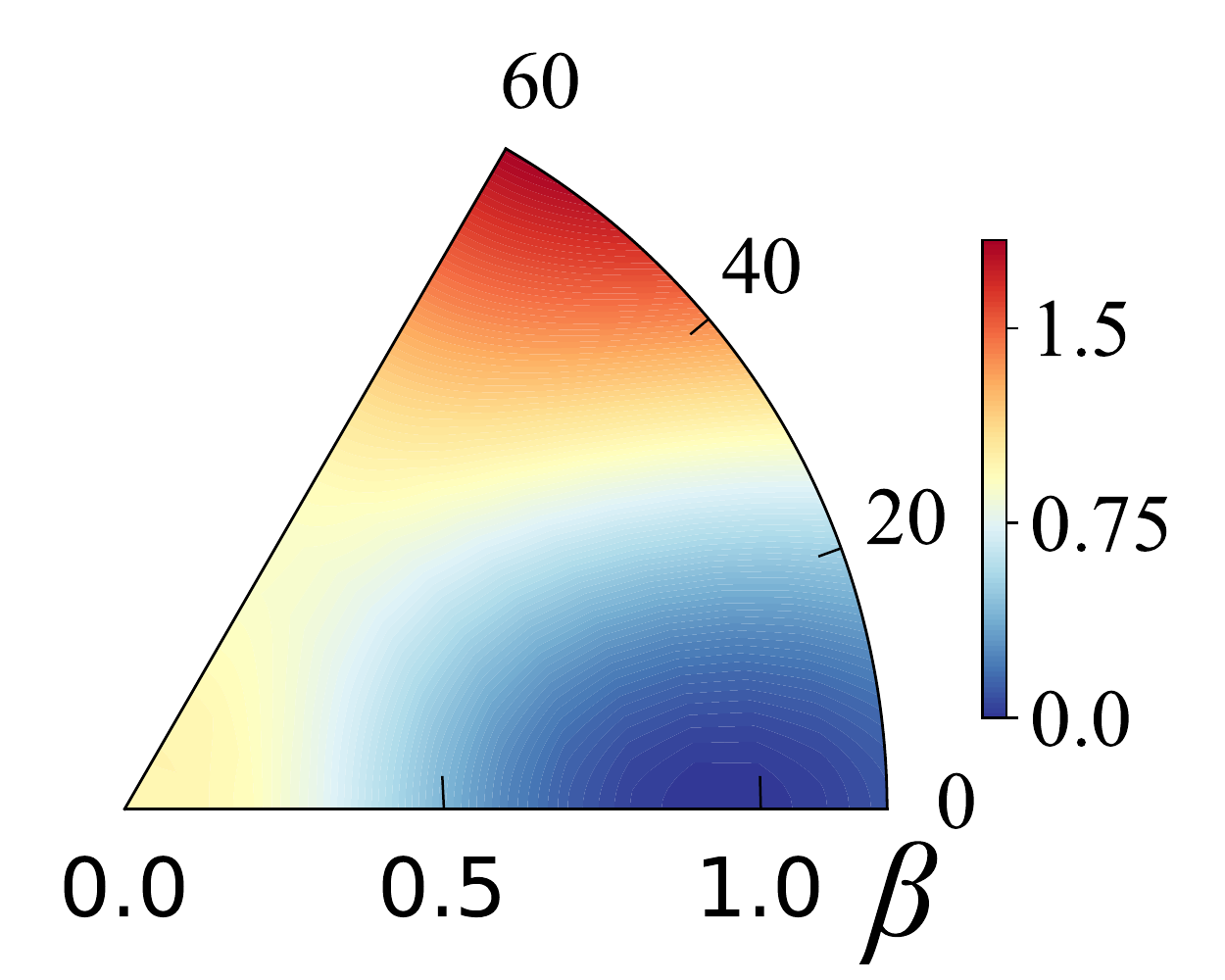}
\put (0,67.5) {\large(f)}
\put (0,50) {\large $^{102}$Zr}
\put (60,70) {$\gamma$(deg)}
\end{overpic}
\begin{overpic}[width=0.19\linewidth]{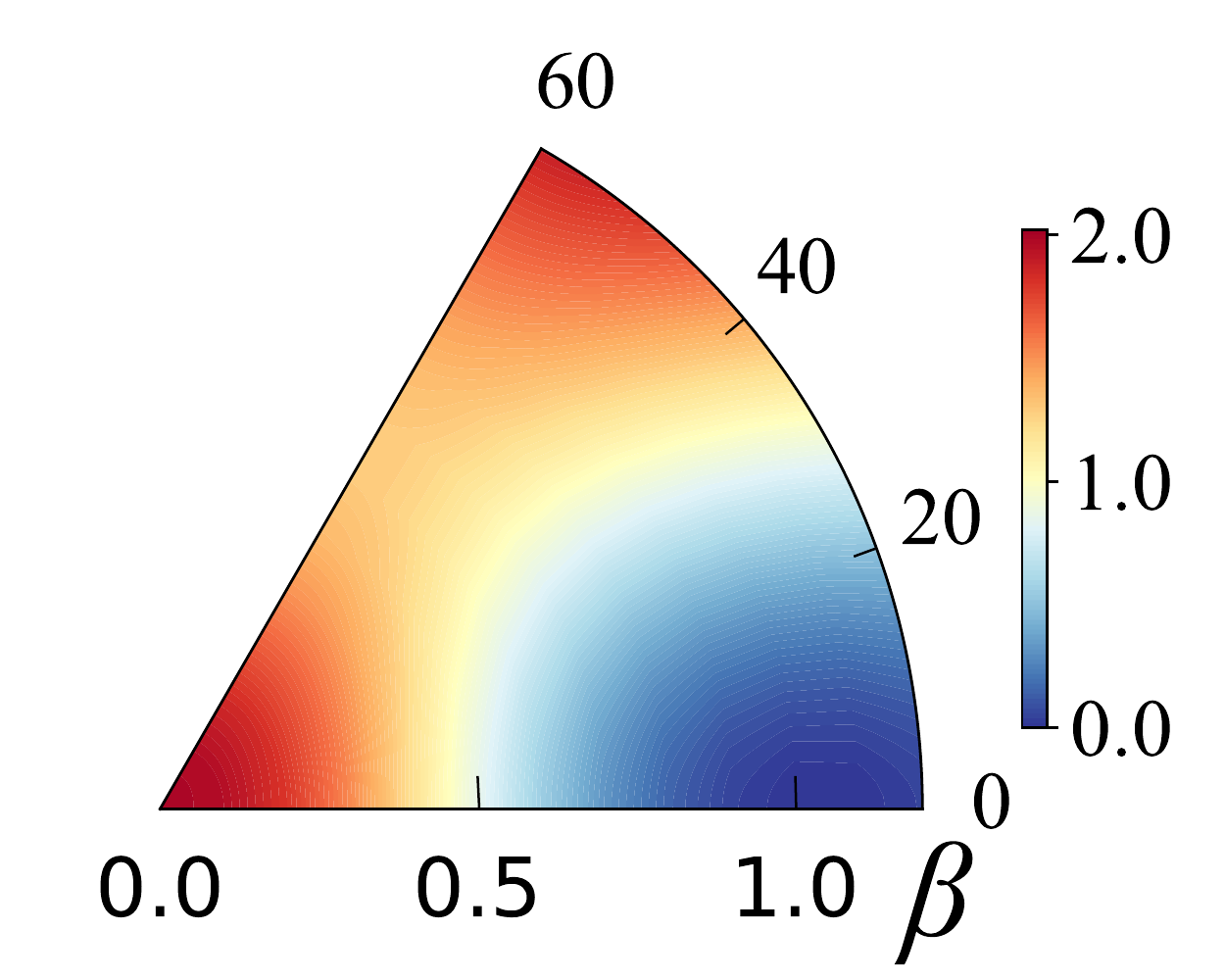}
\put (0,67.5) {\large(g)}
\put (0,50) {\large $^{104}$Zr}
\put (60,70) {$\gamma$(deg)}
\end{overpic}
\begin{overpic}[width=0.19\linewidth]{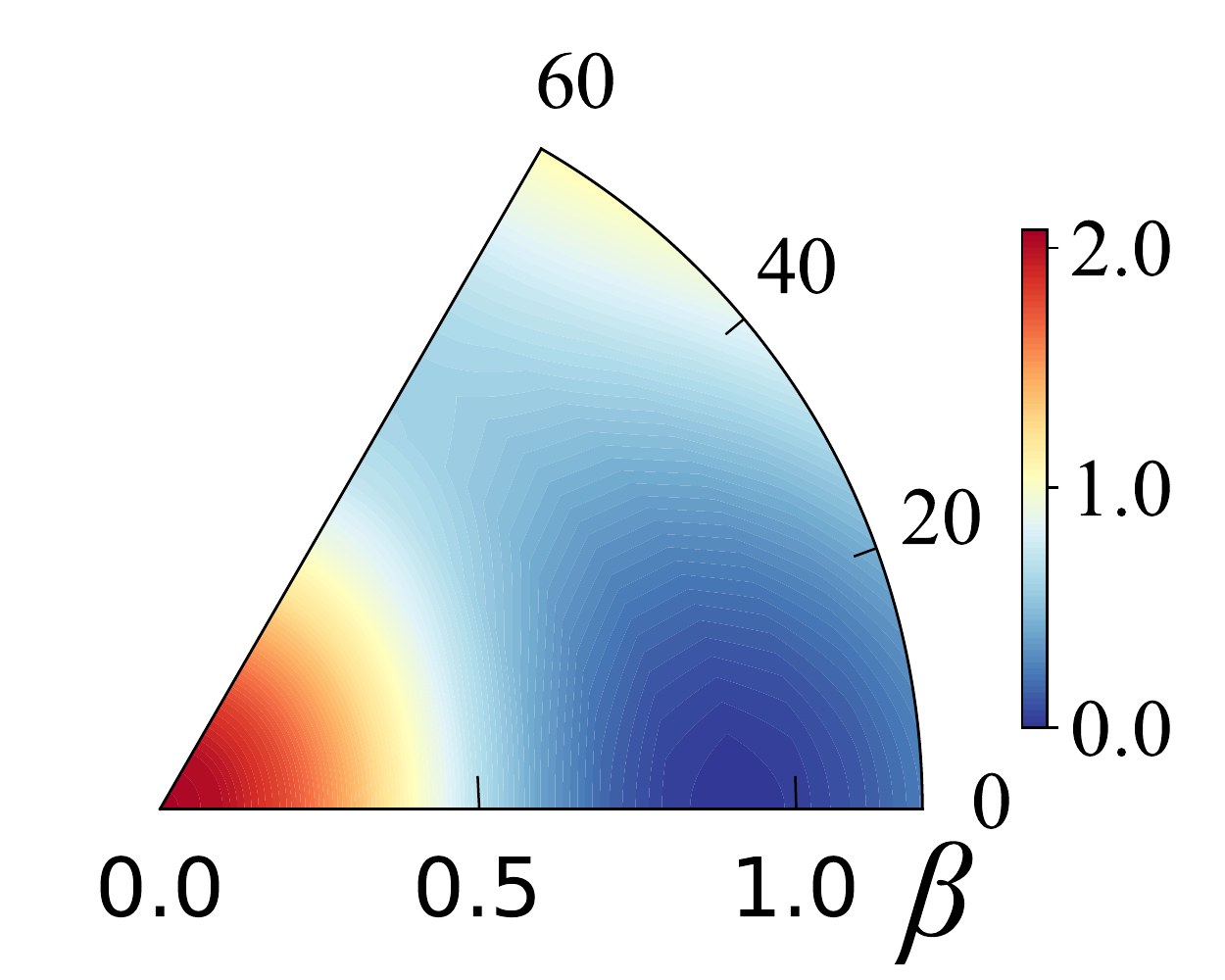}
\put (0,67.5) {\large(h)}
\put (0,50) {\large $^{106}$Zr}
\put (60,70) {$\gamma$(deg)}
\end{overpic}
\begin{overpic}[width=0.19\linewidth]{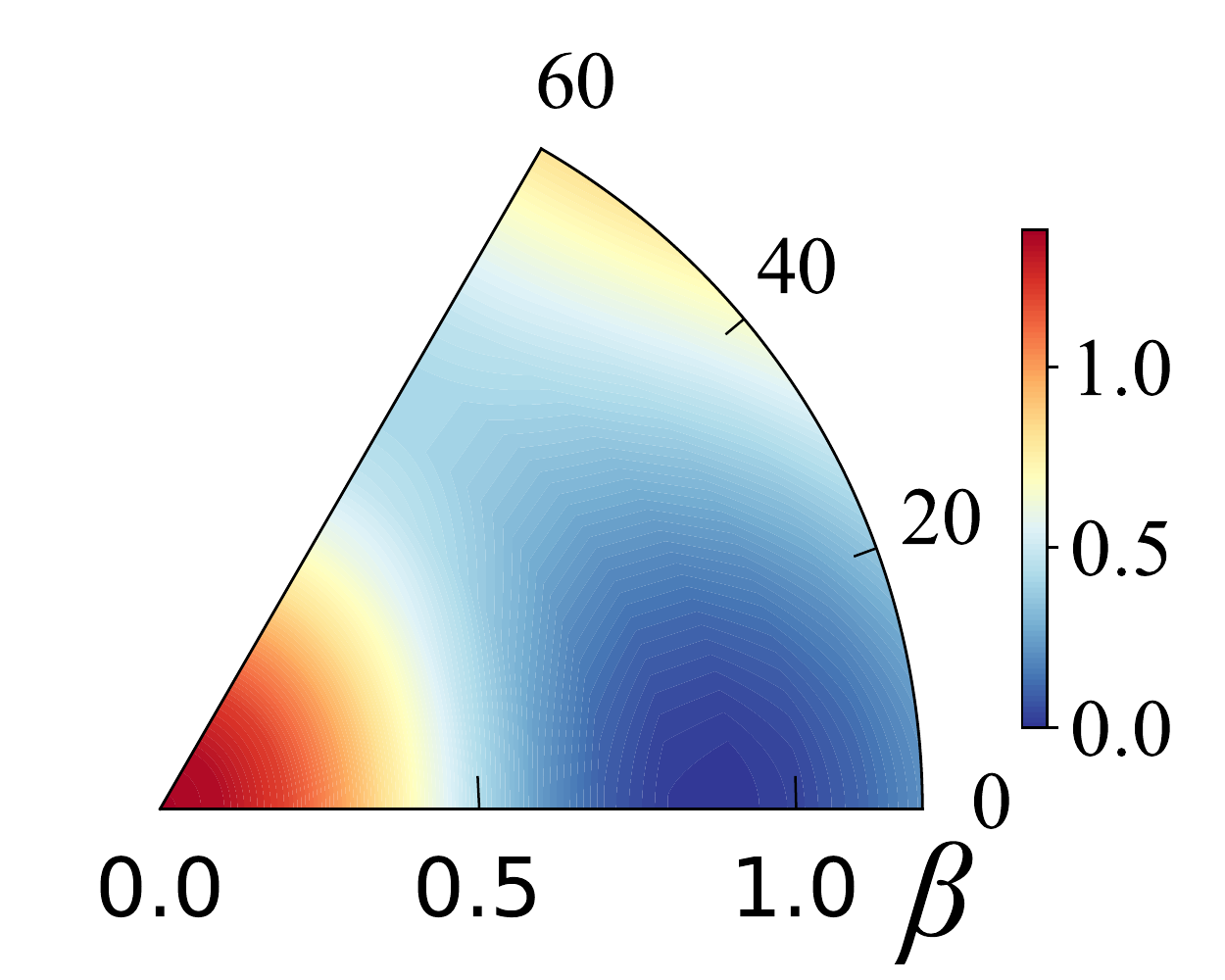}
\put (0,67.5) {\large(i)}
\put (0,50) {\large $^{108}$Zr}
\put (60,70) {$\gamma$(deg)}
\end{overpic}
\begin{overpic}[width=0.19\linewidth]{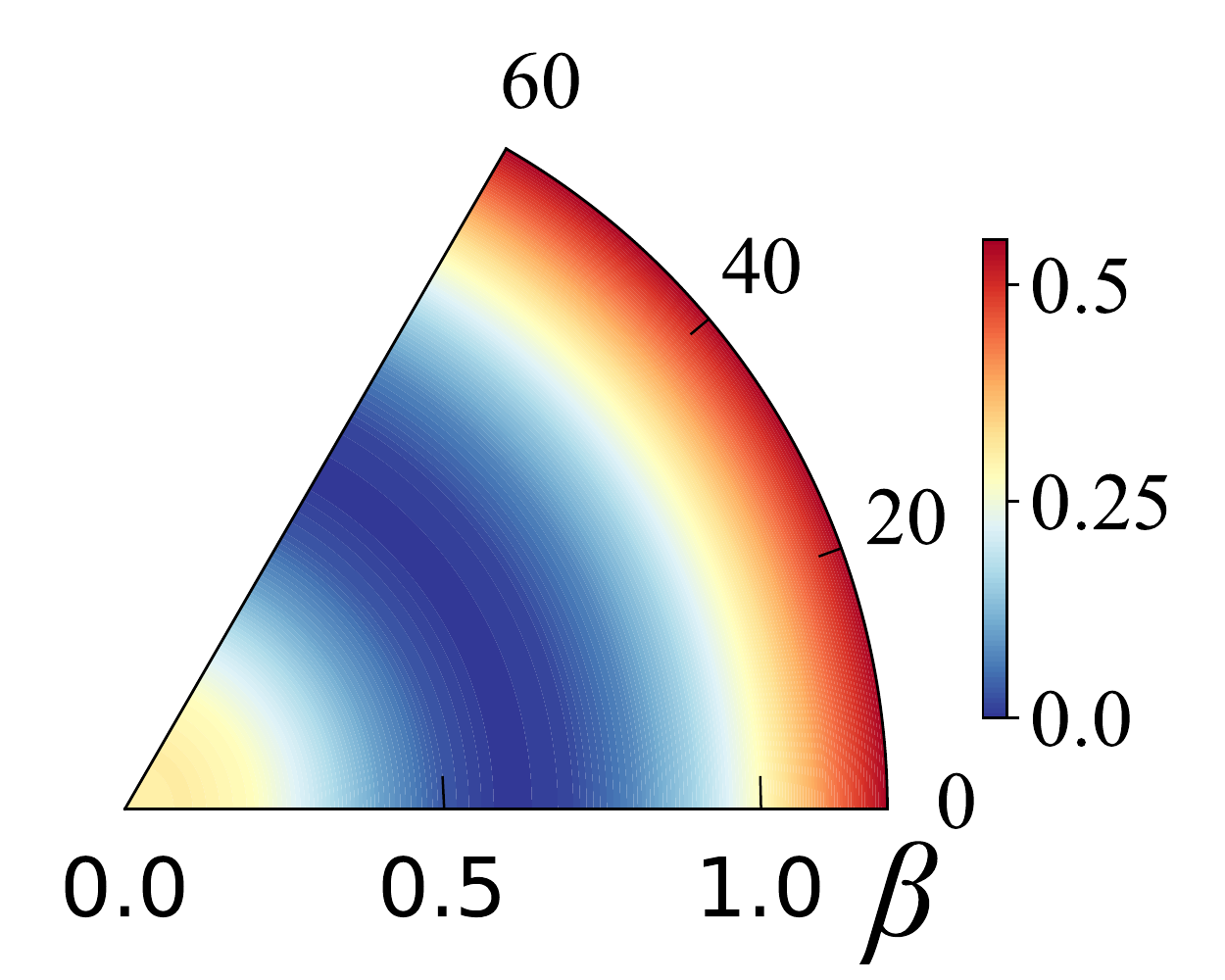}
\put (0,67.5) {\large(j)}
\put (0,50) {\large $^{110}$Zr}
\put (60,70) {$\gamma$(deg)}
\end{overpic}
\caption{\label{Eminus}
\small
Contour plots in the $(\beta ,\gamma )$ plane of the lowest eigen-potential 
surface, $E_{-}(\beta ,\gamma )$, for the $^{92-110}$Zr isotopes.}
\end{figure*}
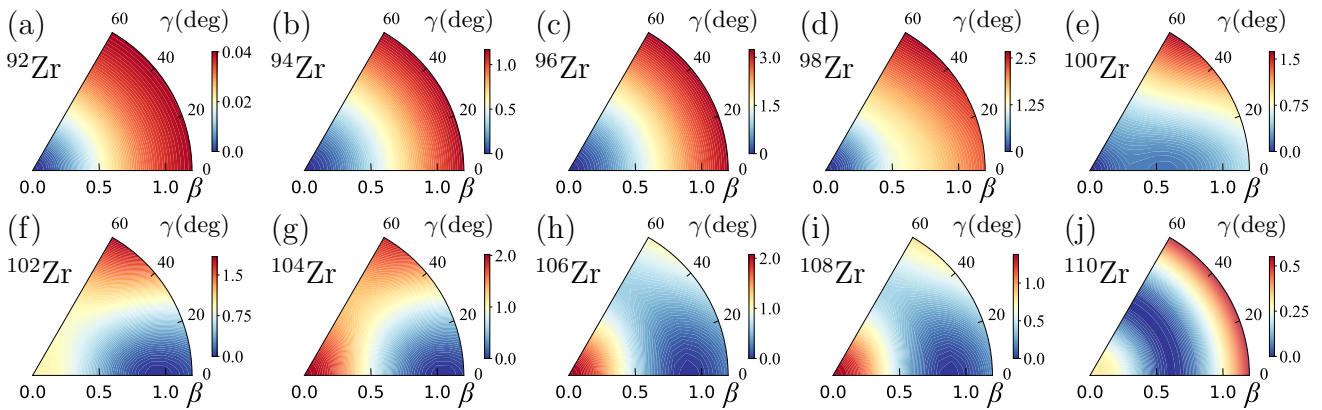

We note that our calculations describe the experimental data in the
entire range $^{92-110}$Zr very well.
A~full account is given in~\cite{gavrielov}. Here we show 
only two examples, $^{100}$Zr and $^{110}$Zr. 
$^{100}$Zr is near the critical point of both Type~I and Type~II QPT 
and yet our description of energy levels and $B(E2)$ 
values is excellent, Fig.~\ref{fig:spectrum}(a)-(b). 
The ground state band, configuration~($B$), 
appears to have features of the so-called X(5) 
symmetry~\cite{iachello-x5}, while 
the spherical configuration~($A$) has now become the excited band $ 0^+_2 $. 
$^{110}$Zr, Fig.~\ref{fig:spectrum}(c)-(d),
appears instead to be an excellent example of SO(6)
symmetry~\cite{arima-so6}, although few experimental data are available.
In general, the results of the current phenomenological study
resemble those obtained in the microscopic approach of the
MCSM~\cite{taka} (which focuses on spectra and $E2$ rates), 
however, there are some noticeable differences. Specifically,  
the replacement $\gamma$-unstable $\rightarrow$ triaxial 
and the inclusion of more than two configurations in the MCSM. 
The spherical state in $^{100}$Zr is identified in the MCSM as $0^{+}_4$,
in contrast to $0^{+}_2$ in the current calculation and the data.
Both calculations show a large jump in 
$B(E2;2^+_1\rightarrow0^+_1)$, between $ ^{98} $Zr and $ ^{100} $Zr, 
typical of a 1$^{st} $ order QPT. This is in contrast with mean-field 
based calculations~\cite{delaroche,mei,nomura16}, 
which due to their character smooth out the phase transitional behavior, 
and show no such jump at the critical point of the QPT
(see Fig.~2 of~\cite{Singh18}).
The observed peak in $B(E2;2^+_1\rightarrow0^+_1)$ for $^{104}$Zr,
is reproduced by the current calculation but not by the MCSM.

As mentioned above, one of the main advantages of the algebraic method is
that one can do both a quantum and a classical analysis. 
In Fig.~\ref{Eminus}, we show the calculated 
lowest eigen-potential $E_{-}(\beta ,\gamma )$. These classical potentials 
confirm the quantum results, as they show a transition from spherical 
($^{92-98}$Zr), Fig.~\ref{Eminus}(a)-(d),
to a flat-bottomed potential at $^{100}$Zr, Fig.~\ref{Eminus}(e),
to axially deformed ($^{102-104}$Zr), Fig.~\ref{Eminus}(f)-(g),
and finally to 
$\gamma $-unstable ($^{106-110}$Zr), Fig.~\ref{Eminus}(h)-(j).

\section{Conclusions}
In this article, we have calculated the spectra and several
other observables for the entire chain of $_{40}$Zr isotopes,
from neutron number 52 to 70,
in the framework of the IBM-CM. The results of the comprehensive
analysis suggest that IQPTs appear to be manifested empirically 
in these isotopes. The latter exhibit a complex phase structure with
two configurations, one spherical~($A$) and the other~($B$)
undergoing first a QPT U(5)-SU(3) and then a crossover SU(3)-SO(6). 
These shape-changing Type~I QPTs occur simultaneously with 
a configuration-changing Type~II QPT, 
in which the normal and intruder configurations cross,
a characteristic pattern of IQPTs.
Further details of our results, including the calculation of spectra and 
transition rates in all the $^{92-110}$Zr isotopes and of other quantities 
not reported here, will be given in a forthcoming publication
based on~\cite{gavrielov}. 
Our method of calculation could also be applied to the $_{38}$Sr isotopes,
which show similar features, and we are planning to do so in a future
publication. The present work provides the first evidence for
intertwined quantum phase transitions in nuclear physics
and may stimulate research for this type of phase transitions in other
fields of physics.\\

\acknowledgements
This work was supported in part by U.S. DOE under Grant No.
DE-FG02-91ER-40608 and by the US-Israel Binational Science Foundation 
Grant No. 2016032. We thank R.~F. Casten for fruitful discussions.
Upon completion of this work, we have
learned from J.~E.~Garc\'\i a-Ramos, of another ongoing calculation
for the Zr isotopes in the IBM-CM framework.

\end{document}